# FOMO: Topics versus documents in legal eDiscovery

Herbert L. Roitblat[†]
Mimecast
Ventura, CA USA
herb@eroitblat.com

## ABSTRACT

In the United States, the parties to a lawsuit are required to search through their electronically stored information to find documents that are relevant to the specific case and produce them to their opposing party. Negotiations over the scope of these searches often reflect a fear that something will be missed (Fear of Missing Out: FOMO). A Recall level of 80%, for example, means that 20% of the relevant documents will be left unproduced. This paper makes the argument that eDiscovery is the process of identifying responsive information, not identifying documents. Documents are the carriers of the information; they are not the direct targets of the process. A given document may contain one or more topics or factoids and a factoid may appear in more than one document. The coupon collector's problem, Heaps law, and other analyses provide ways to model the problem of finding information from among documents. In eDiscovery, however, the parties do not know how many factoids there might be in a collection or their probabilities. This paper describes a simple model that estimates the confidence that a fact will be omitted from the produced set (the identified set), while being contained in the missed set. Two data sets are then analyzed, a small set involving microaggressions and larger set involving classification of web pages. Both show that it is possible to discover at least one example of each available topic within a relatively small number of documents, meaning the further effort will not return additional novel information. The smaller data set is also used to investigate whether the non-random order of searching for responsive documents commonly used in eDiscovery (called continuous active learning) affects the distribution of topics—it does not.

## 1 Introduction

The parties in a US lawsuit have an obligation to perform a reasonable search for relevant information, and this search process can be highly burdensome and expensive. The parties in many cases negotiate search methods and success criteria and these negotiations are often disputatious. One party or the other is often concerned that something important might be missed if the search is less than perfect. That fear of missing out (FOMO) contributes to the high cost of eDiscovery.

If a producing party reported that their eDiscovery search process found 80% of the responsive documents (80% Recall), that means that 80% of the responsive documents have been identified, and that 20% of the responsive documents have not. The fundamental question addressed by this paper is whether the missing documents justify the burdensome effort to try to increase the proportion of responsive documents that are identified and produced.

I address this question by modelling and by comparison with the actual categorization of content in two data sets.

### 1.1 Background

An organization involved in a lawsuit may have thousands or even millions of documents that must be evaluated for relevance to a specific case (the responsive documents). A multibillion-dollar industry has emerged to help these organizations and their law firms discharge their search obligations. The kind of information that is sought, the locations and time period that must be searched, and the methods for finding it, are often negotiated by the parties near the start of the litigation. The party requesting the information, which is typically the plaintiff (the party bringing the action) has an interest in making sure that the search is as complete as possible. The producing party is usually interested in producing as little as necessary and in minimizing the cost of the search process. Not every case fits this pattern specifically, but enough of them do to make the contribution of this paper important to the parties and the courts.

The US Federal Rules of Civil Procedure (available at https://www.law.cornell.edu/rules/frcp), which govern the operation of US Federal courts, specify that (Rule 1) the rules "should be construed, administered, and employed by the court and the parties to secure the just, speedy, and inexpensive determination of every action and proceeding." They also specify (Rule 26(g)) that the parties must employ a reasonable discovery effort, not a perfect one. The key word here is "reasonable," and the parties and the court may find it difficult to achieve a consensus on just how much work constitutes a reasonable effort (see also Oot, Kershaw, & Roitblat, 2010). This judgment can benefit from a systematic analysis of how likely it is that there is significant information to be found among the unproduced relevant documents.

The primary question concerns the likelihood that some topic will be missed in the set of identified responsive documents but could be found with further effort. If further effort is unlikely to produce new information, then that effort may be considered unreasonable (ceteris paribus). Many attorneys have anecdotes of the time when they did actually find a critical piece of information after further effort, but until now there has been no effort to estimate how likely such events are. Low probability events do happen, but by definition, they are unlikely.

In this paper, I refer interchangeably to topics, categories, and facts. Each of these terms carries connotative baggage, so it may be better to refer to them each as factoids. A unique piece of information (factoid) may consist of a high-level category or a specific fact with truth values (for example, to texts that discuss the concept of a fraud or to those that provide specific evidence of a fraudulent act).

For modeling purposes, all of these terms, as used here, refer to units of information that can be assigned a probability of occurrence. A document consists of one or more factoids and each factoid can appear in more than one document. This model is not concerned with how those factoids are defined or to how the documents or factoids are identified.

In order for a factoid to be omitted from the identified set, it would have to be low enough prevalence to not be contained there. In order for it to be found in the missed set, it must be high enough prevalence to be contained in that set.

Typically, in eDiscovery, the parties know the size of the document set that is produced and the Recall level. That is, they know the number of responsive documents that have been identified and they estimate the number of responsive documents remaining to be identified. For example, the producing party may start with a million documents. After a search, they may identify 100,000 of these documents as being relevant to the case. They may also estimate (based on a validation set of documents or some other method) that these identified documents constitute 80% of all of the relevant documents (80% Recall). Therefore, some 25,000 relevant documents remain to be identified. Whether further effort is justified depends on a number of factors, among them is the likelihood of identifying significant new information—factoids—in the set of missed documents.

## 2   Related work

Fundamentally, we are interested in the probability that a factoid will be omitted from the documents in the identified set and contained in the relevant documents in the missed set. There are several closely related problems.

The coupon-collector's problem is to estimate the number of draws needed to select a complete set of objects randomly from a probability distribution sampled with replacement. It is named after the problem of collecting coupons from, say boxes of breakfast cereal or trading cards from packets of gum, where each packet contains one coupon/ticket/baseball card (Flajolet, Gardy, & Thimonier, 1992). This problem has found broad application in computing and elsewhere (Boneh, & Hofri, 1997).

The eDiscovery factoids correspond to the coupons and the documents correspond to the packets. Unlike the standard coupon collector's problem, though, the coupons in eDiscovery differ in their probability of occurrence, which makes the standard analysis intractable (Ferrante & Saltalamacchia, 2014; see also Brown, Peköz, & Ross, 2008; Shioda, 2007).

This problem can be viewed as a bins and balls occupancy problem. How many balls need to be randomly thrown into bins until each bin has at least one ball? Like the coupon-collector's problem, analysis of this problem is greatly complicated by unequal probabilities of landing in each bin, but there are some suggestions of simpler computations for upper and lower bounds (Atsonios, Beaumont, Hanusse, & Kim, 2011). But this analysis relies on the probabilities of each coupon to be selected from a power-law distribution, but even if they are so distributed, we do not know its parameters.

Another, similar problem is the cover time for a graph (e.g., Ding, Lee, & Perez, 2011). How long does it take for a random walk on a graph to visit every vertex at least once? The graph in this case would be the topics and the connections between them would be the sequence in which the factoids were encountered. How many time steps would be needed to touch each topic at least once?

Heap's law (Ferrante & Frigo, 2012/2018) is an empirical law that describes the proportion of unique words that are used in a text of a given size. It is closely related to Zipf's law (or the Zipf-Mandelbrot distribution), which concerns the rank order of the terms used in a text and their proportion of usage in the text. Heaps law, more generally, concerns the size of a random sample needed to observe a specific number of different records. If each factoid is a record in this sense, then Heaps law could be used to estimate the number of documents that would be needed to find those topics.

For these approaches to be used successfully in eDiscovery, the attorneys would have to know the number of topics to be discovered and their probabilities. There is usually no way for them to know these things, so we need to take a simplified approach. When the topics differ substantially in frequency, then the number of documents that need to be searched to find at least one example of each topic will depend strongly on the topic or topics with the lowest prevalence. That is the approach taken here for eDiscovery. By comparison, I show what happens when we do know the topics and their prevalence to support the validity of using the rarest topic as a guide.

## 3 Modeling

For simplicity, we assume that each document has exactly one topic. We will later relax that assumption. If we assume a certain confidence level (*c*, say 95%), then we can estimate the maximum probability that a missed document would have from Equation 1.

$$ln(1-p) = \frac{ln(1-c)}{n_1} \quad (1)$$

Here p is the top of the 95% confidence interval for the prevalence of a topic that was not contained in the identified set. Each document in the identified set provides an opportunity for the occurrence of a factoid, so the probability that a factoid will be omitted is the probability that it is not contained in the first document, and not contained in second document, …, and not contained in the final document.

In the missed set, k is the confidence of finding a fact with probability *p* from Equation 1, and $n_2$ documents, the number of documents in the missed set. Note that $n_2$ is the number of responsive documents that have not been identified. We may know the number of these documents, but in eDiscovery, we do not know their identity. They will be mixed in with a (usually large) number of non-relevant documents.

$$k = 1 - (1-p)^{n_2} \quad (2)$$

Following the example given earlier, if the identified set contained 100,000 responsive documents, the top of the 95% confidence interval for the prevalence of an omitted document would be about 0.00003 (prevalence of 0.003%, or 1 per 33,381 responsive documents). The confidence that a topic with this probability will contained in at least one document in the missed set is about 52.71%. Therefore, the confidence that a topic will be omitted from the identified set and contained in the missed set is the approximated as the product of the confidence that it will not be in the identified set (1.0 – confidence level) and the confidence that will occur in the missed set yielding 2.63%. In other words, in about 2.63% of cases or fewer will there be at least one additional topic among the missed set of documents.

At lower Recall levels, we are more likely to discover a new fact in the missed set, but this increase is rather small. At 70% Recall, the probability of finding a new fact increases to 3.6%. At 60% Recall, it increases to 4.3% of cases.

## 4 Microaggression and LDA

Section 3 described some general theory behind the identification of topics in eDiscovery. In the second part we apply that thinking to some actual communications.

A basic idea behind topic modeling is that documents consist of words that are derived from some mixture of topics. The topics can be viewed usefully as latent variables. In this view, each document is generated by a selection from a distribution of topics and then by a selection from a distribution of words for each topic. Each topic is associated with a distribution of words.

In this study, I used latent Dirichlet allocation (Blei, Ng, & Jordan, 2003) to extract the set of topics from all of the known relevant documents from a collection. I then trained a machine learning categorizer to identify relevant documents (without knowledge of the topics assigned to them) in order to work backwards and determine which of the topics was covered by the documents in the identified set and which were covered by the documents that were missed.

### 4.1 The data

The documents for this study consist of a set of micro-aggression and distractor messages collected by Breitfeller et al. (2019). Microaggressions are statements or actions, that express a prejudiced attitude toward a member of a marginalized minority.

Breitfeller and colleagues mounted a substantial effort to collect and identify microaggressive statements from social media. They started with self-reported incidents of microaggressions and then developed methods of annotating further examples. See their paper for details. Because they are categorized (as microaggressions or not), these messages provide a good set of relevant (microaggressions) and nonrelevant documents for this investigation.

A few example microaggressive statements:
- At least I don't sit on my ass all day collecting welfare, I EARN my money.
- Did you get this job because you're pregnant?
- Do your parents make sacrifices so that you can go to our school?

I selected all 1,676 microaggressive statements as the to-be-identified set against a haphazardly selected set of 2,000 other texts from their collection. The difference between microagressions and other statements can be quite subtle, so this task is challenging for any text classifier.

### 4.2 Identifying topics

Several methods (e.g., latent semantic analysis, e.g., Deerwester et al., 1990, probabilistic latent semantic analysis, Hofman, 1999) are available for inferring the topics associated with each document. For this study, I chose, as mentioned, latent Dirichlet allocation (Blei, Ng, & Jordan, 2003), using the Scikit Learn function (Pedregosa, et al., 2011). These methods require the researcher to specify the number of expected topics.

I chose 100 categories for the Latent Dirichlet analysis. With 1,676 short documents to work with, specifying a hundred topics seems like a very large number, with an average of about 16.8 documents per topic. For simplicity, I focused on the one topic per document that had the highest score. The actual number of documents associated with each topic varied widely (See Figure 1) The most common topic was contained in 3.28% of the microaggression messages, the least common topic was represented in 0.24% of the microaggression messages.

By way of example, the 15 most closely associated words for the first four Microaggression topics are shown in Table 1.

**Table 1. The top words for the first four topics identified using Latent Dirichlet Analysis**

| Topic 0 | tell, mixed, person, ask, race, knew, woman, mom, want, feel, white, oh, black, know, excellent |
|---|---|
| Topic 1 | student, female, street, deserve, sorry, male, sir, miss, people, know, american, black, african, difference, theres |
| Topic 2 | exotic, walk, meaning, harder, means, gets, place, muslim, want, country, students, woman, thought, native, american |
| Topic 3 | life, know, english, woman, sisters, friend, right, white, really, older, teacher, language, oh, youre, born |

The number of aggressive statements associated with each topic are shown in Figure 1 and Figure 2.

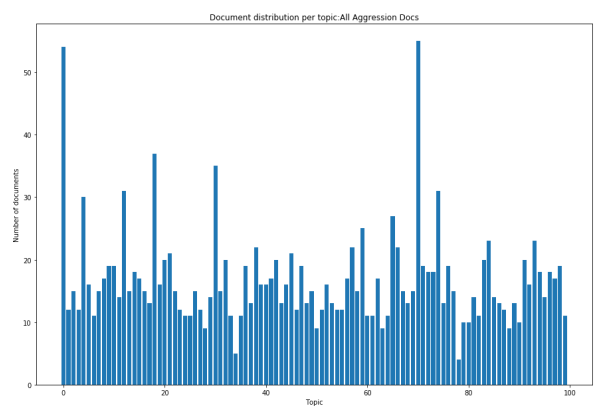

**Figure 1. The distribution of microaggression statements per each of 100 topics for all documents.**

These are the topics associated with the full collection of aggressive statements.

Figure 2 shows the categories sorted in descending order of the number of documents. The number of documents per topic corresponds to a power-law (Zipf) distribution (R2 of the log/log regression = 0.90).

### 4.3 Categorization learning

It is common in eDiscovery to use some form of supervised machine learning to identify the responsive or relevant documents in a collection. In this section I used a naïve Bayes classifier (MultinomialNB) to distinguish between the relevant (microaggressive) and the non-relevant (non-aggressive) classes and train_test_split to divide the examples into training and testing groups.

A randomly selected 1,257 microaggression and 1,500 non-relevant statements were used for training and the rest were used for testing (the hold-out set). No topic information was used during this phase of the study.

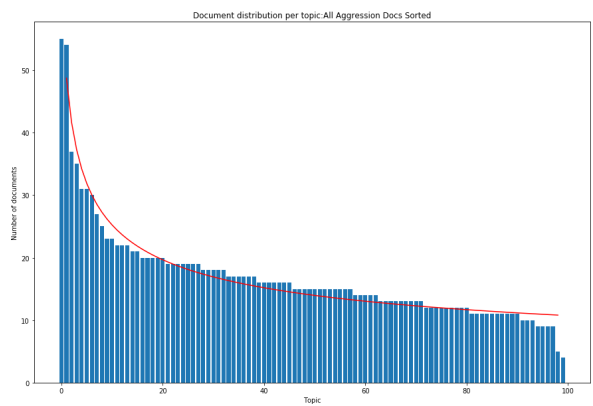

**Figure 2. The number of documents for each topic sorted in descending order. The red line shows the expected number of documents under the predicted power-law distribution.**

The initial quality of the classifier was assessed against the hold-out set. The classifier achieved 75% Recall and 85% Precision on the hold-out set. When the relevant documents from the holdout set were added to the original training results (as is typical for eDiscovery, all of the identified relevant documents need to be produced, except for those deemed privileged) the overall produced-set Recall was 81%. In eDiscovery, the concern is not with how well a learning model will generalize to future data, but with how well it will accurately classify the documents in the instant collection. A total of 318 documents were misclassified as non-relevant (the missed set).

Figure 3 shows the distribution of topics for the hits. The number of documents associated with each topic varied substantially.

All 100 topics were represented among the identified set messages (counting only the presence of the strongest topic for each message). These messages identified all of the available topics even though the machine learning process failed to identify 318 documents (19% of the responsive ones). No topics were missed among the identified set, so no topics were available to be recognized only in the missed set.

Because all 100 of the topics were found among the identified set, no additional topics would be gained from attempting to search for the documents that remained after the Naïve Bayes achieved 81% Recall.

In this situation, we happen to know each of the topics and their probabilities, but in eDiscovery we would not have this information. From this analysis, there is confidence that topics with higher probability would be discovered, but the confidence interval for the topic probabilities includes 0.0, meaning that there may be no novel topics to be recovered from further effort. In most cases, there not be any.

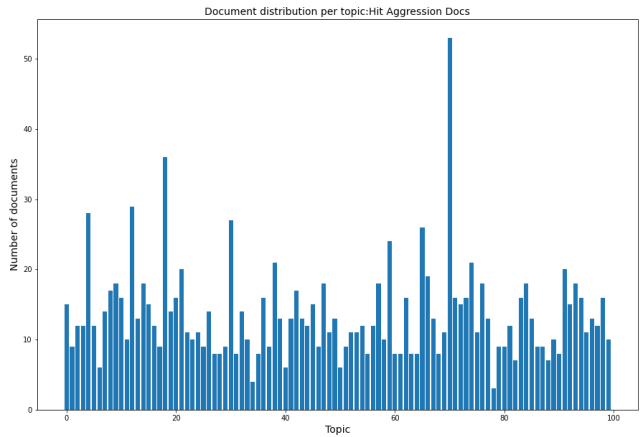

**Figure 3. The number of documents holding each topic from among those that were classified as relevant by the Naïve Bayes Classifier. Only the documents in the identified set are shown.**

Although no claim is made in this model for how the probabilities are distributed or for the order in which they are discovered, the idea of using the top of the confidence interval for the probability of unseen topics assumes that documents are encountered randomly. But eDiscovery is not a random search process, so it makes sense to consider whether these results depend on the randomness of the selection process. A biased search process, for example, could more exhaustively select some topics for identification and neglect other topics, leaving them to the missed set.

### 4.5 Nonrandom document selection

For example, search in eDiscovery often uses a document selection method called Continuous Active Learning (Cormack and Grossman, 2014). This approach presents documents to a team of reviewers (labelers) in an order determined by the method's classifier (usually a support vector machine; Cortes and Vapnik 1995) so that the documents predicted to be most likely to be relevant are presented first. After each training batch (for example, a hundred documents in this study), the system updates the model and makes new predictions. It then predicts a new batch derived from the remaining documents, which is again sorted into order of descending probability of relevance. These are the documents presented for review. As a result, continuous active learning may present items for review in runs of similar documents and thus prefer some topics over others. On the other hand, continuous active learning does not have access to the factoid information about a document, only to the words in the documents. As a result, the ordering is likely to be only weakly related to the order in which the factoids are encountered.

To evaluate the argument that these results depend strongly on random selection of training documents, I repeated the same observations using a simulation of continuous active learning.

I replaced the Naïve Bayes classifier with a Support Vector Machine Classifier (again from Scikit), similar to the method used by Cormack and Grossman. Following their methods, training of the classifier began with a small number of randomly selected messages and their labels (microaggression or not). After the training on the first batch, each additional batch presented for labeling the 100 documents most likely to be responsive as judged by the current state of the classifier. Then, the classifier was updated and another batch was generated for judgment. The labels for each provided document were selected by a simulated oracle, using the original labels provided by Breitfeller et al. No effort was made in either phase of this study to emulate the human error that may occur when labeling documents.

The batch presentation and update process was continued until the system achieved a Recall level (81.3%) comparable to that achieved using the Naïve Bayes system (81.0%).

To assess the degree to which continuous active learning labeled documents in a less random way than was used with Naïve Bayes classifier, I computed the Kolmogorov-Sinai entropy (K-S entropy; Sinai, 2009) for the sequence of topics from document to document in the two learning regimes. For the Bayesian classifier, where training occurred in random order, the K-S entropy was 58.72 bits. For continuous active learning, K-S entropy was slightly lower at 55.87 bits. The sequence of topics was slightly more predictable for continuous active learning than for Bayesian classifier, but this this small difference was inconsequential in the present context.

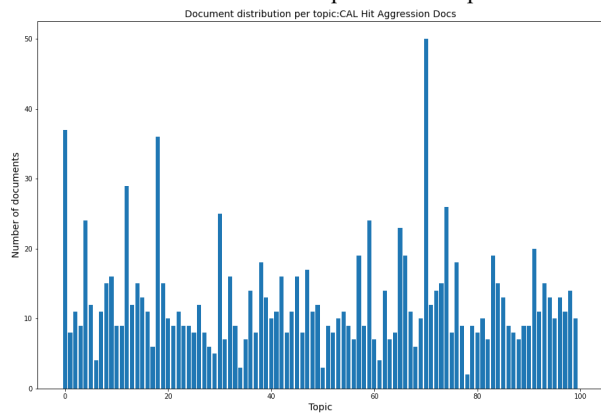

**Figure 4. The number of documents holding each topic among the identified set resulting from continuous active learning and a support vector machine.**

Even though the documents chosen for training in this simulation were not selected randomly, these results also show that all 100 topics were again identified among the documents designated by the machine learning system as relevant. No topics were left to be discovered among the missed documents. There is, thus, no evidence that non-random document selection has any significant effect on topic coverage among the identified relevant documents.

The same results were found using a Naïve Bayes Classifier trained on a random selection of documents as with a simulation of a currently popular eDiscovery method, continuous active learning. In both cases, all of the available topics were discovered among the documents predicted to be responsive. The ordered selection of documents for labeling (and including in the training set) had no discernible effect on the measure of interest, whether the documents predicted to be responsive were sufficient to identify the available topics. It is possible that a more sensitive test might eventually find some differences resulting from the search method, but so far, we have no evidence that using a non-random method invalidates the conclusions of this analysis.

In retrospect, it is perfectly reasonable to expect similar topic coverage for different methods of training document ordering. Two methods that achieve 80% overlap with the set of relevant documents must have substantial overlap with one another. Both the naïve Bayes classifier and the continuous active learning systems correlated to the same level (about 80% Recall) with the set of all relevant documents, so they would be expected to overlap in about 64% or more of those documents, and that would be enough to cover all of the topics.

## 5 Web page categorization

Another sample of documents and their labels was obtained in the ordinary course of my company's business. This sample consisted of almost 12 million web pages, each of which was categorized by a commercial system into one or more of 64 categories. I did not have access to the web pages, themselves, but only to the categories assigned to each one. We could still examine the number of pages that would have to be examined in order to find examples of these 64 categories.

As in the previous section, we can know the relative prevalence of documents with each tag. Unlike the previous part, the documents in this section can each contain more than one topic, averaging 1.37 topics per page.

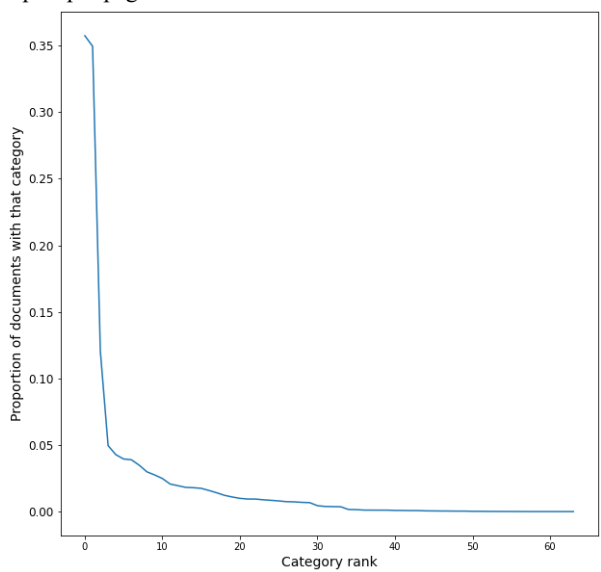

**Figure 5. The relative frequency of each topic category ordered by rank.**

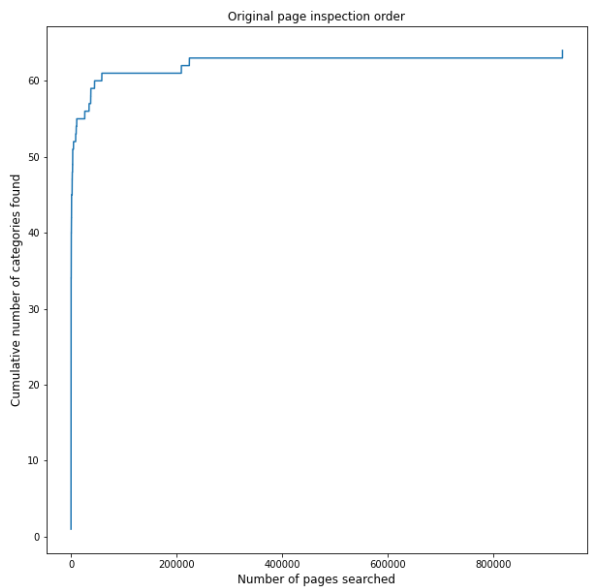

**Figure 6. The cumulative number of categories observed as documents were scanned.**

The topics varied widely in the number of documents that contained them, and thus in their probability. The most common topic appeared in about 36% of the documents (about 1 in 3). The rarest topic occurred in about 0.00014% (about 1 in 71,429 documents). See Figure 5.

Figure 6 shows the cumulative number of categories that were observed as additional documents were scanned in the ordinary course of business. These documents were examined as they were requested by users, so their order was driven by the users' information needs, not some random number generator. Most of the topics were observed during the first (relatively) few documents, the steep part of the curve in Figure 6. All of them were observed within 931,226 documents.

In the ordinary course of business, the rarest topic was encountered after only about 8% of the documents had been categorized, corresponding to 8% Recall. That is, the classification had stopped there, it would have identified 100% of the topics, but only 8% of the relevant documents in this set. The other 92% would not have been identified.

But one series does not allow us to draw any statistical inferences concerning the probable number of documents that would have to be examined. For that, I ran some simulations, which consisted of randomly shuffling the pages and counting the number of documents that would have to be categorized to find all 64 topics. Each shuffle would correspond to a single case, because any individual case would have only one document ordering (as in the ordinary course of business). Each shuffle was evaluated until at least one example of each topic was found.

I repeated this shuffling 2,000 times. Figure 7 shows the number of shuffles in which the corresponding number of documents were categorized to find all 64 topics. The maximum number of documents needed was 5,124,553 over all 2,000 shuffles (as shown by the last bar).

Table 2 shows the number of documents that were needed to find all of the topics, and the corresponding proportion of all of the labeled pages. This proportion corresponds to the minimum level of Recall to identify all topics. In 10% of the shuffles, 301,028 or fewer documents were needed. Because there were almost 12 million categorized documents, we can say that in 10% of the shuffles, we needed only 2.51% of these relevant documents to be identified to cover all topics. In 95% of the shuffles, we needed no more than 18.4% of the documents to identify all 64 topics. These results mean that we would be 95% confident that we could find all of the topics if we identified only 18.4% if the relevant documents in this collection. Higher Recall rates would return more documents, but not more topics (because here we know that there are no topics left to discover).

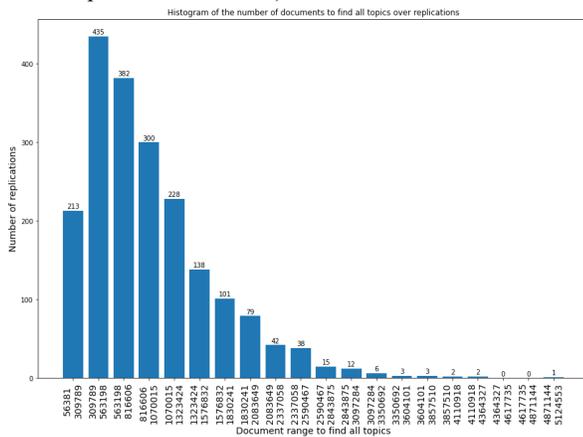

**Figure 7.** Histogram of the number of documents that needed to be searched to find all 64 topics after a random shuffle.

The model described in Section 2 suggests that when we do not know the exact number of categories or their probability, we can use the likelihood of the rarest topics to estimate the likelihood of missing a topic in the identified set and finding it in the missed set. Based solely on the rarest topic, we can predict that we would need an average of 704,180 documents if each page contained only a single topic. The mean count over 2,000 shuffles to find all 64 topics was 933,456.1 pages. In any given shuffle, the number of pages needed to hit all 64 topics would be the maximum of the time needed to reach each of the topics, so it would necessarily require more documents than the number needed to reach any one (see Table 3).

All 64 topics were collected when the final outstanding topic (the terminal topic) was encountered. The final outstanding topic was not always the rarest one, because this is a stochastic sample. If the time to first encounter a topic is $x_i$, then the time to encounter all of the topics is $\max(x_i)$, and $x_i$ is a random variable. Topic 58, then was the terminal topic only when all of the other topics had already been observed. Moreover, unlike the model, each page could be labeled with more than one topic, but the co-occurrence pattern was also complicated, particularly for these rare topics that served alternately as the terminal topics.

**Table 2. The number of documents were scanned in order to find all 64 topics.**

| PERCENTILE | NUMBER OF DOCUMENTS | RECALL LEVEL |
|---|---|---|
| 10 | 301,028 | 2.51% |
| 20 | 413,766 | 3.46% |
| 30 | 529,128 | 4.42% |
| 40 | 667,325 | 5.57% |
| 50 | 797,179 | 6.66% |
| 60 | 957,359 | 8.00% |
| 70 | 1,149,967 | 9.61% |
| 80 | 1,371,954 | 11.46% |
| 90 | 1,843,165 | 15.40% |
| 95 | 2,202,935 | 18.40% |

**Table 3. The last topics found that completed the set.**

| TOPIC # | N | P | PREDICT | MEAN MAX |
|---|---|---|---|---|
| 63 | 930 | 0.0000014 | 704,180.24 | 1,033,235.33 |
| 62 | 874 | 0.0000016 | 630,056.00 | 956,115.17 |
| 61 | 80 | 0.0000066 | 151,532.46 | 415,454.10 |
| 60 | 45 | 0.0000098 | 102,316.79 | 330,199.51 |
| 58 | 39 | 0.0000113 | 88,674.55 | 318,253.74 |
| 59 | 32 | 0.0000105 | 95,008.44 | 307,858.91 |

Among the 2,000 shuffles, five different topics were the terminal topic. Table 3 shows the number of times the set was completed by each topic, the topic's probability and the predicted number of documents that would be needed on each shuffle based solely on the topic's prevalence. The last column shows the mean number of pages required to complete the set for each terminal topic, which means that all other topics were found on that simulation in fewer pages.

Equation 1 predicts that the maximum probability of a missed topic in the identified set, based on the mean size of that set, The observed mean was 933,456, which would be 0.0000032 with a confidence level 0f 0.95.

## 6 Discussion

I think that the key eDiscovery insight in the present project is the idea that documents are carriers of information. The Federal Rules of Civil Procedure discuss "Electronically Stored Information," not documents. Rule 26 provides "Specific Limitations on Electronically Stored Information. A party need not provide discovery of electronically stored information from sources that the party identifies as not reasonably accessible because of undue burden or cost." And further, "the court must limit the frequency or extent of discovery otherwise allowed by these rules or by local rule if it determines that: the discovery sought is unreasonably cumulative or duplicative, or can be obtained from some other source that is more convenient, less burdensome, or less expensive." The process described in this paper provides a means for assessing the degree to which electronically stored information is likely to be cumulative or duplicative based on the volume of documents produced and level of Recall involved.

The number of documents that it takes to achieve full coverage of a topic set can be approximated by the prevalence of the least probable topics. In this case, the least likely category had a very low prevalence, occurring only once in 704,180 documents. This is exceedingly rare in my experience. The chance of being struck by lightning is around 1 in 500,000. The chance of being dealt a royal flush in poker is 1 in 649,740. Finding this topic in any given document is more likely than getting hit by a meteor, but less likely than being hit by lightning.

Any actual eDiscovery exercise may have more or fewer responsive documents, higher or lower Recall, and more or fewer topics/factoids. The model suggests the likelihood that further investigation will yield at least one new topic. Exceedingly rare events do happen, but their occurrence is more dependent on chance than on any systematic effort.

The FOMO approach is not a form of eDiscovery, rather it is an analysis that describes how topics might be allocated between the documents that are produced and the documents that are missed. It does not distinguish among search methods. It does not directly solve the problems of how to search, which custodians to search, or how to assess the relevance of the documents. It does not eliminate the need for a good-faith reasonable discovery process.

The FOMO analysis is silent as to what the factoids are or how they are characterized. It is concerned with the probability of encountering one or more examples of each topic, not with the content of a specific topic or its probative or dispositive value. The topics in these presented examples are fairly broad. They are the sort that might be useful to a business enterprise. They are roughly equivalent to the issue tags that are commonly used in eDiscovery. In a particular case, these high-level topics may be broken down into finer grained categories. The number of these subtopics is neither known nor critically important, but their prevalence is. Some documents may be more probative than others, but these subsets could be treated as different topics.

This analysis does not speak to how to identify topics or how to compare documents to determine whether they contain the same topic. Although there are machine-learning methods that can be helpful in making these comparisons, they are outside the scope of this analysis.

The FOMO model only concerns itself with the relevant documents. The ones that are identified as part of the discovery process are relatively easy to assess because they have been identified. In actual practice, however, the documents in the missed set are unknown. The probabilities given in the FOMO model are predictions of how likely a new topic is to be encountered among the responsive documents that remain to be identified. The prevalence is the probability of a given responsive document having this topic, not the prevalence of the topic among the whole set of unlabeled documents. Many relevant documents will have to be examined to find a new topic, that has escaped notice by the search process that was used. That discovery effort is outside the scope of the FOMO modeling approach, but it could be quite extensive. The responsive documents are typically a small subset of the total number of documents and an even smaller subset of the remainder after the identified responsive ones have been identified. Finding the relevant ones in the context of all of the irrelevant could be extremely expensive and burdensome, potentially equal to the original discovery.

In one of the first studies of the effectiveness of using computerized search in eDiscovery, Blair and Maron (1985) worked with attorneys on keyword searching. "The information-request and query formulation procedures were considered complete only when the lawyer stated in writing that he or she was satisfied with the search results for that particular query (i.e., in his or her judgment, more than 75 percent of the 'vital,' 'satisfactory,' and 'marginally relevant' documents had been retrieved)." Interestingly, when Blair and Maron eventually evaluated the success of these searches, they found that the lawyers' queries had actually resulted in 20%, not 75% Recall, as the lawyers had estimated and that they said was necessary.

The Blair and Maron study is often cited as an example of the poor accuracy of keyword searching, but it is also an indicator that lower levels of Recall may be sufficient to provide all of the necessary information in a case. Their lawyers were satisfied that 20% Recall met their information need.

This view is entirely consistent with the results shown in Table 2. Even fairly low levels of Recall can return all of the available topics, under some circumstances. In this example, 18.4% Recall was sufficient in 95% of the cases to return all of the available topics.

Roitblat, Kershaw, and Oot (2010) compared two human review teams and two machine learning systems intended to identify responsive documents. The two human teams achieved 49% and 54% Recall respectively, and the two machine systems achieved 46% and 53% Recall. Interestingly, however, the human reviewers only agreed on 28% of the documents that either team found responsive. The two teams were about equally effective at identifying responsive documents (they showed 49% and 54% Recall), but, importantly in this context, they disagreed about which those documents were. Presumably, both groups identified the same (or at least similar) information, but this information was contained in different documents.

The conclusion of the FOMO eDiscovery analysis is that a reasonable discovery process is unlikely to leave new relevant information behind in the unidentified missed set of documents.

In eDiscovery we generally do not know the number, the content, or the prevalence of examples of each factoid. There are some indications that whatever topics there are, their prevalence is distributed as an approximation to Zipf's law (see Figure 2). More specific facts are also likely to be more rare, so, although, they may exist, the rarer they are, the less likely they are to be encountered by any specific search method, making the relevant documents in the missed set, even more difficult to identify.

A common question from lawyers and judges asks: The analysis says that there is at least one example of a topic among the identified documents, but what if there is a more probative example of that topic? Don't we have to search for that?

This analysis suggests that such a search may not be necessary or fruitful. The highly probative examples of any general topic and the general topic itself, could be considered separate factoids. They would each have a specifiable (even if unknown) prevalence among the responsive documents. If that prevalence is high enough, there is no reason to think that they would be more likely to be found in the missed set when the identified set is typically significantly larger (at Recall levels above 50%). There is nothing in highly probative topic examples that should make them systematically more difficult to identify. Even if they are more difficult to identify, they would be likely to be missed everywhere and not be found in either identified nor the missed sets.

However, it is true that the missed responsive documents have not yet been identified when considering whether additional effort is necessary. The producing party would first have to identify these documents to be able to produce the topics that are within them. The court and/or the parties have to decide whether that effort can be justified. In making that decision, it might be helpful to think of it as a gamble: How much would you pay for the opportunity to search for a lottery ticket that might or might not pay off (there could be no new topics, no payoff), and if it does pay off would pay some amount between zero dollars (the topic is new, but does not advance the case) and the full value of the case (a so-called "smoking gun")? Aside from all of the legal decision factors that must be balanced, valuing this wager is itself a difficult question, but it illustrates the exact problem that must be solved to determine if the effort is reasonable.

Lawyers may see the focus on topics as the appropriate unit of discovery as a radical departure from ordinary practice. However, it addresses directly the concept of whether there is likely to be new and useful information in documents that are predicted to exist (because Recall is estimated to be less than 100%) but not yet identified after a reasonable search. The analyses presented here suggest that there is likely little new information to be gained from the disproportionate effort needed to search for additional relevant documents among those that were missed by the initial discovery search.

Nothing here says that there cannot be some new topic or some fact that was not already identified in an actual eDiscovery exercise, but it does say that such new topics are unlikely. These studies have not proved there are no unicorns to be found in the missed set, but finding one will be unlikely, closer to wishful thinking than likely principled outcome.

## References

Adler, I., Oren, S., & Ross, S.M. (2003). The coupon-collector's problem revisited. Journal of Applied Probability. 40(2): 513–518. https://adler.ieor.berkeley.edu/ilans_pubs/coupons_2003.pdf

Ioannis Atsonios, Olivier Beaumont, Nicolas Hanusse, and Yusik Kim. 2011. On power-law distributed balls in bins and its applications to view size estimation. In Proceedings of the 22nd international conference on Algorithms and Computation (ISAAC'11). Springer-Verlag, Berlin, Heidelberg, 504–513. DOI:https://doi.org/10.1007/978-3-642-25591-5_52

David C. Blair and M. E. Maron. 1985. An evaluation of retrieval effectiveness for a full-text document-retrieval system. Commun. ACM 28, 3 (March 1985), 289–299. DOI:https://doi.org/10.1145/3166.3197

David M. Blei, Andrew Y. Ng, and Michael I. Jordan. 2003. Latent Dirichlet allocation. J. Mach. Learn. Res. 3, null (3/1/2003), 993–1022.

A. Boneh, and M. Hofri. "The coupon-collector problem revisited — a survey of engineering problems and computational methods." Stochastic Models 13 (1997): 39-66.

Luke Breitfeller, Emily Ahn, & David Jurgens, & Yulia Tsvetkov,. (2019). Finding Microaggressions in the Wild: A Case for Locating Elusive Phenomena in Social Media Posts. 1664-1674. 10.18653/v1/D19-1176.

Mark Brown, Erol a. Peköz, and Sheldon m. Ross. 2008. Coupon collecting. Probab. Eng. Inf. Sci. 22, 2 (March 2008), 221–229. DOI:https://doi.org/10.1017/S0269964808000132

Gordon F. Cormack and Maura F. Grossman. 2014. Evaluation of machine-learning protocols for technology-assisted review in electronic discovery. Proceedings of the SIGIR Conference 2014 (2014), 153–162. https://doi.org/10.1145/2600428.2609601.

Corinna Cortes and Vladimir Vapnik, 1995."SupportVector Networks," Machine Learning, 20, 273-297.

Scott Deerwester, Susan T. Dumais, George W. Firmas, Thomas K. Landauer, and Richard Harshman. Indexing by latent semantic analysis. Journal of the American Society for Information Science, 41(6):391-407, September 1990.

Jian Ding, James R. Lee, and Yuval Peres. 2011. Cover times, blanket times, and majorizing measures. In Proceedings of the forty-third annual ACM symposium on Theory of computing (STOC '11). Association for Computing Machinery, New York, NY, USA, 61–70. DOI:https://doi.org/10.1145/1993636.1993646

Doumas, Aristides V. and V. Papanicolaou. "The logarithmic Zipf law in a general urn problem." Esaim: Probability and Statistics 24 (2020): 275-293.

Marco Ferrante & Monica Saltalamacchia (2014). The Coupon Collector's Problem. MATerials MATemàtics, 2014 (No. 2). 35 pp. Publicació electrònica de divulgació del Departament de Matemàtiques de la Universitat Autònoma de Barcelona. http://mat.uab.cat/matmat_antiga/PDFv2014/v2014n02.pdf

P. Flajolet, D. Gardy, L. Thimonier, Birthday paradox, coupon collectors, caching algorithms and self-organizing search, Discrete Appl. Math., 39:207-229, 1992.

Thomas Hofmann. 1999. Probabilistic latent semantic indexing. In Proceedings of the 22nd annual international ACM SIGIR conference on Research and development in information retrieval (SIGIR '99). Association for Computing Machinery, New York, NY, USA, 50–57. DOI:https://doi.org/10.1145/312624.312649

Marco Ferrante and Nadia Frigo (2012/2018). On the expected number of different records in a random sample. arXiv:1209.4592v1 [math.PR]

Patrick Oot, Anne Kershaw, & Herbert L. Roitblat 2010. Mandating reasonableness in a reasonable inquiry. Denver University law review 87(2)

Fabian Pedregosa, Gaël Varoquaux, Alexandre Gramfort, Vincent Michel, Bertrand Thirion, Olivier Grisel, Mathieu Blondel, Peter Prettenhofer, Ron Weiss, Vincent Dubourg, Jake Vanderplas, Alexandre Passos, David Cournapeau, Matthieu Brucher, Matthieu Perrot, and Édouard Duchesnay. 2011. Scikit-learn: Machine Learning in Python. J. Mach. Learn. Res 12, 2825–2830.

Herbert L. Roitblat, 2019. FOMO and eDiscovery. https://www.edrm.net/wp-content/uploads/2019/05/FOMO-and-eDiscovery-Herb-Roitblat-Ph.D..pdf

Vassilis G. Papanicolaou, George E. Kokolakis, and Shahar Boneh. 1998. Asymptotics for the random coupon collector problem. J. Comput. Appl. Math. 93, 2 (July 13, 1998), 95–105. DOI:https://doi.org/10.1016/S0377-0427(98)00058-2

Yakov Sinai (2009), Kolmogorov-Sinai entropy. Scholarpedia, 4(3):2034.

S. Shioda. 2007. Some upper and lower bounds on the coupon collector problem. J. Comput. Appl. Math. 200, 1 (March, 2007), 154–167. DOI:https://doi.org/10.1016/j.cam.2005.12.011